# The low temperature thermal expansion of single-walled carbon nanotube bundles saturated with nitrogen

A.V. Dolbin[1], V.B. Esel'son[1], V.G. Gavrilko[1], V.G. Manzhelii[1], S.N. Popov[1],

N.A.Vinnikov[1], B. Sundqvist[2].

[1] Institute for Low Temperature Physics & Engineering NASU, Kharkov 61103, Ukraine

[2] Department of Physics, Umea University, SE - 901 87 Umea, Sweden

## Abstract

The effect of a $N_2$ impurity on the radial thermal expansion coefficient $\alpha_r$ of single-walled carbon nanotube bundles has been investigated in the temperature interval 2.2 – 43 K by the dilatometric method. Saturation of nanotube bundles with $N_2$ caused a sharp increase in the positive magnitudes of $\alpha_r$ in the whole range of temperatures used and a very high and wide <u>maximum</u> in the thermal expansion coefficient $\alpha_r$ (T) at T ~ 28 K. The low temperature desorption of the impurity from the $N_2$-saturated powder of bundles of single-walled carbon nanotubes with open and closed ends has been investigated.

## Introduction

Carbon nanotubes have been attracting an ever growing interest due to their obvious potential and considerable promise for technological applications (e.g., molecular electronics, gas-filled detectors, nanoemitters, etc.) Nevertheless, some properties of carbon nanotubes (CNTs) and their bundles, among them the low temperature thermal expansion, still remain obscure. Dilatometric measurements of the radial thermal expansion coefficient $\alpha_r$ on single-walled carbon nanotube (SWNT) bundles at T = 2—120 K [1-2] show that $\alpha_r$ is positive above 5.5 K and negative at lower temperatures. The negative $\alpha_r$ values can be attributed to the low-frequency vibrations of individual nanotubes at liquid helium temperatures. The low frequency part of the vibrational spectrum of individual CNTs is described by a negative Grüneisen coefficient, which is typical of two-dimensional systems [3]. At higher temperatures the expansion of SWNT bundles make a dominant contribution to the thermal expansion of the system which is described by a positive Grüneisen coefficient, as in three-dimensional systems [3]. As was found previously, saturating SWNT bundles with Xe [2] and $H_2$ [4] impurities causes a drastic increase in the radial thermal expansion coefficient (rTEC) and causes local maxima in the dependence $\alpha_r(T)$. The sharp growth of $\alpha_r$ is most likely associated with the effect of the introduced gas molecules on the transverse vibrations of the system. The $\alpha_r(T)$ peaks may be due to the redistribution of the impurity molecules in the grooves and at the surface of the SWNT bundles. A controlled desorption of the impurities from the doped samples showed that the $\alpha_r(T)$ peaks decreased with the impurity concentration. It was interesting to extend the investigations mentioned [1,2,4] and to find out how the radial thermal expansion of SWNT bundles could be influenced by gases consisting of nonspherical molecules. In this study the radial thermal expansion of $N_2$-saturated SWNT bundles was measured in the temperature interval 2.2—43 K. To interpret the results obtained, we needed information about the concentration and the spatial distribution of the $N_2$ molecules in SWNT bundles. For this purpose, the temperature dependence of $N_2$ desorption from the $N_2$-saturated bundles of SWNT with closed and open ends was investigated through a controlled desorption procedure.

# 1. Low temperature desorption of N₂ impurity from CNT powder

## Experimental technique

The $N_2$ desorption from a CNT powder was investigated in the temperature interval 46—120 K using a special cryogenic device. Its design and the experimental technique are detailed elsewhere [2]. According to the manufacturer, the starting CNT powder (CCVD method, Cheap Tubes, USA) contained over 90% of SWNTs, the average outer diameter of the tubes being 1.1 nm. Two samples were used: the starting CNT powder and a CNT powder after an oxidative treatment applied to open the CNT ends (see [2] for details). The samples of closed and open CNTs were saturated with $N_2$ through the same procedure. Prior to measurement each sample was evacuated at $10^{-3}$ Torr for 72 hours directly in the measuring cell to remove possible gas impurities. The cell with the sample was then filled with $N_2$ to the pressure 760 Torr and cooled to T = 46 K. The saturation gas was 99.997% pure $N_2$ with $O_2 \leq 0.003\%$ as an impurity. Doping the nanotubes with $N_2$ was continued at T = 46 K. $N_2$ was added in small portions as soon as the previous one was adsorbed. The variations of the $N_2$ pressure were registered in the course of sorption of each portion. Nitrogen was fed until the pressure in the cell reached an equilibrium value of 0.1 Torr, slightly below the equilibrium pressure of $N_2$ at 46 K (0.46 Torr [5]). The sample was then cooled down to 43 K and the gas not adsorbed by the CNT powder was removed from the cell. Thereafter, the $N_2$-saturated CNT powder was heated in steps of 3 K and the $N_2$ desorption from the CNT was investigated by measuring the desorbed gas quantities.

## Results and discussion

The temperature dependence of the gas quantities released on heating the sample (mole per mole of the CNT powder, i.e. the number of impurity molecules per carbon atom) is illustrated in Fig. 1.

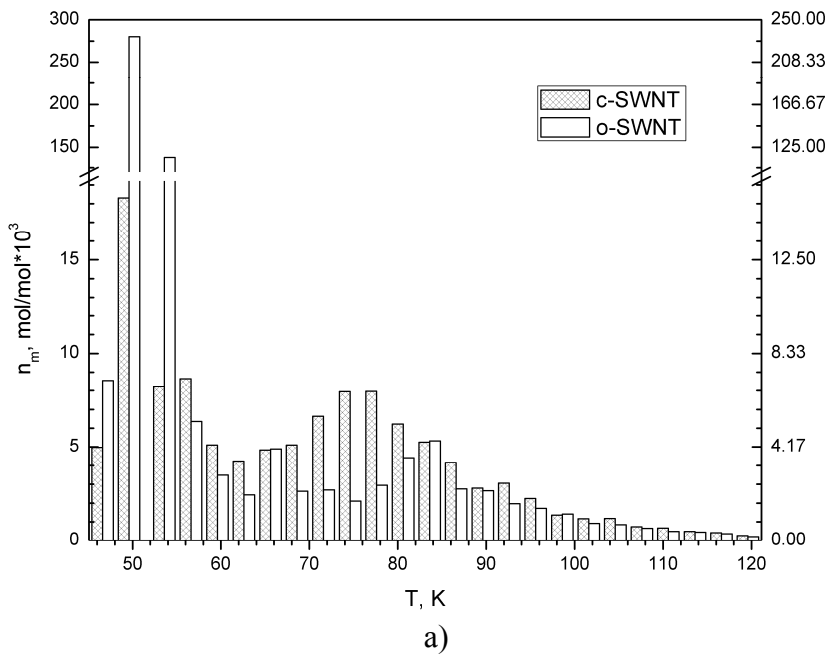

a)

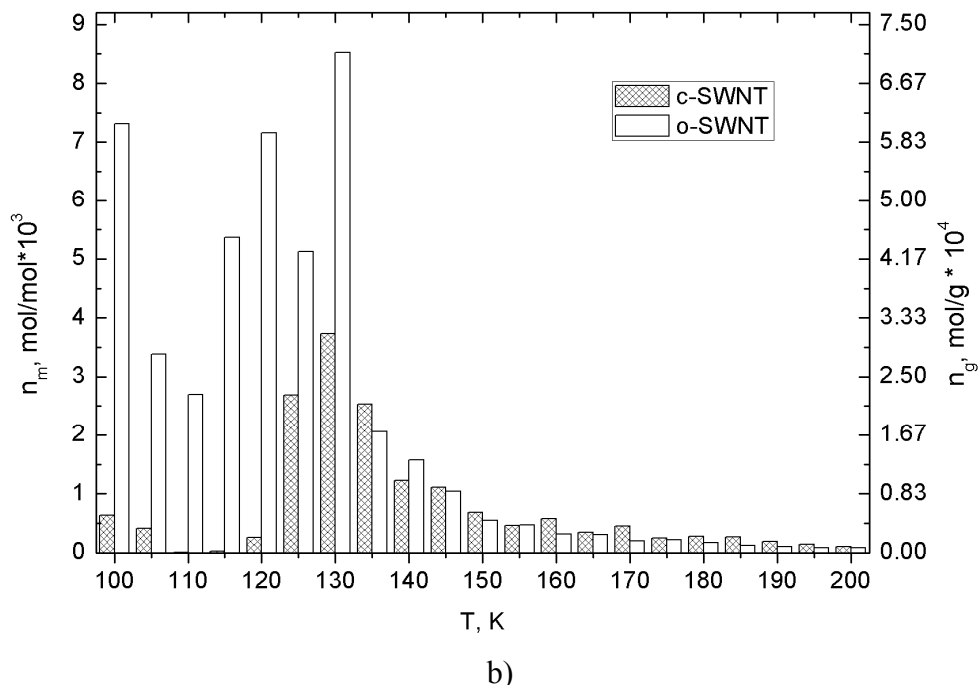

b)

Fig. 1. Temperature distribution of $N_2$ (a) and Xe (b) impurities desorbed from the powder of c-SWNTs (dark columns) and o-SWNTs (light columns).

Note that the $N_2$ quantities desorbed from the sample and adsorbed on saturation were equal within the experimental error, which indicates an almost complete removal of the $N_2$ impurity from the sample on heating. The sites of possible sorption of $N_2$ molecules in a bundle of infinite equal-diameter CNTs are shown schematically in Fig. 2.

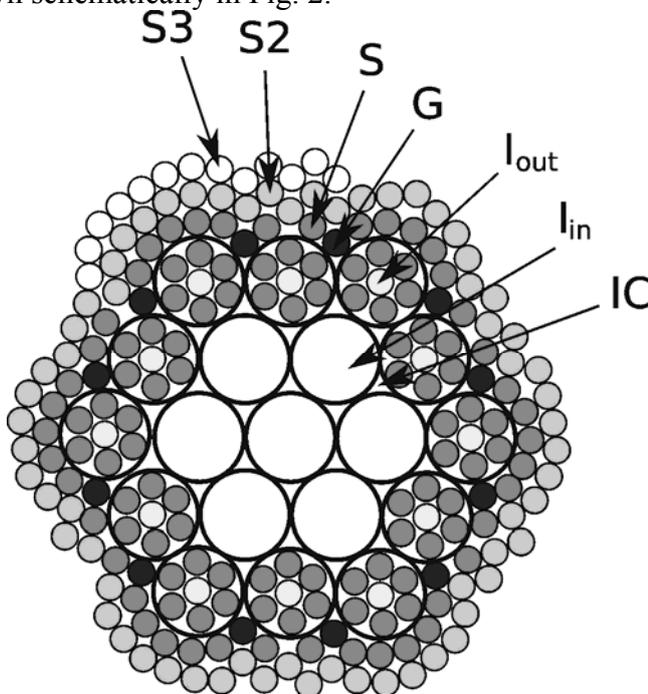

Fig. 2. Sites of possible $N_2$ sorption in a bundle of infinite equal-diameter SWNTs with lateral surface defects:

$I_{in}$ — interior sites in the voids of the CNTs forming the central part of the bundle;

$I_{out}$ — interior sites in the voids of the CNTs forming the outer surface of the bundle;

IC — interstitial channels;

G — grooves between the neighboring tubes at the bundle surface;

S — side surface of the bundle;

S2 — the second layer of $N_2$ molecules at the bundle surface;

S3 — the third layer of $N_2$ molecules at the bundle surface.

The total quantities of $N_2$ desorbed from the starting and oxidated SWNT powders are given in the table along with the corresponding data for desorbed Xe and $H_2$.

Table. The total quantities of $N_2$, Xe [2], and $H_2$ [4] desorbed from CNTs (mole/mole %, mass %).

| Impurity | c-SWNT | | o-SWNT | |
|---|---|---|---|---|
| | mol/mol, % | mass % | mol/mol, % | mass % |
| $H_2$ | 10.0 | 1.67 | 8.07 | 1.35 |
| Xe | 1.64 | 7.38 | 4.71 | 21.2 |
| $N_2$ | 11.2 | 26.1 | 46.4 | 108 |

The high concentration of $N_2$ sorbed by c-SWNTs can be explained as follows. $N_2$ molecules can occupy the grooves at the surface of SWNT bundles (G in Fig. 2) and form two or three layers at the bundle surface (S, S2, S3 in Fig. 2). They can also penetrate through inevitable surface defects inside the nanotubes that form the outer surface of the bundle. Besides, some nanotubes within a c-SWNT bundle have open ends and can also be filled with $N_2$. Here we disregard the sorption of $N_2$ by the impurities available in the sample (fullerenes, amorphous carbon, cobalt, etc.).

The diagrams of $N_2$ and Xe desorption exhibit two distinct peaks (see Fig. 1) observed for both the starting and oxidated SWNT powders. It was found [6-8] that in the process of sorption the $N_2$ molecules forming a two-dimensional layer at the bundle surface and a quasi-three-dimensional phase in the interior CNT voids had the weakest interaction with the SWNT bundles (in contrast to Xe atoms having the highest binding energy in the CNT interiors [9]). The peak at ~50 K in the desorption diagram corresponds to the removal of $N_2$ from the sites of low binding energy, while near 75 K most likely accounts for the removal of more strongly bound $N_2$ molecules that form a one-dimensional phase in the grooves at the bundle surface or are localized inside the bundles in the relatively large interstitial channels between different-diameter tubes [10].

It is found in [2] that the air-oxidative treatment applied to open the CNT ends enhanced the Xe sorption. However, our comparative experiments on saturating o-SWNT and c-SWNT bundles with $H_2$ [4] showed that the c-SWNTs absorbed more hydrogen than the oxidated ones. This was attributed to surface defects (mainly vacancies in the carbon structure) present in the CNTs of the starting powder before the oxidative treatment. Small (as compared to Xe atoms) $H_2$ molecules could penetrate easily into the interstitial channels (IC) and inside the CNTs not only through their opened ends but also along the ICs through the vacancies in the CNT walls [11]. The oxidates produced by the oxidative treatment at the vacancy sites blocked the interstitial channels [12] and thus diminished the total quantity of $H_2$ sorbed by the CNTs. The gas-kinetic diameter σ of the $N_2$ molecule ($\sigma_{N2} = 3.708$Å) is much closer to the σ of the Xe atom ($\sigma_{Xe} = 3.921$ Å) than to that of the $H_2$ molecule ($\sigma_{H2} = 2.96$ Å) [13]. The transverse dimensions of most of the interstitial channels are thus prohibitive for $N_2$ molecules [14], and sorption is permitted only through the relatively large channels (not specified in Fig. 2) formed between different-diameter nanotubes within one bundle [10]. One might expect that the oxidative treatment would enhance the $N_2$ sorption in the CNTs. Indeed, the investigation of the sorptive capacity of the oxidated CNTs showed about a 4.1 time increase in the $N_2$ quantity absorbed by the SWNT sample (see the Table). The sharp increase in the $N_2$ quantity desorbed from the oxidated SWNT powder was observed in the temperature interval 49.6—53.6 K (Fig. 1a). This may occur because, firstly, the oxidative treatment caused some "disintegration" of the SWNT bundles [15] and thus increased the areas accessible for sorption at the outer bundle surface whose energy of binding to the impurity molecules is lower than that in the grooves at the CNT surface [6]. Secondly, the oxidative treatment enables the $N_2$ molecules to penetrate through the opened ends of the CNTs into their internal voids where they form a quasi-three-dimensional phase. The binding energy of the $N_2$ molecules occupying the CNT interiors is rather low (see above) and comparable to that of the $N_2$ molecules that form two-dimensional layers at the outer surface of the bundle [8]. This accounts for the sharp growth of the desorption peak at T ≈ 50 K for the oxidated SWNT powder.

There is a qualitative difference between the desorption diagrams obtained for the SWNT bundles saturated with the $N_2$ and Xe impurities. After opening the ends of the CNTs, the adsorption peaks increased in both the low and high temperature regions for the Xe-doped samples and only in the low temperature region for the $N_2$-SWNT bundle system (see Fig. 1a,b). The difference is due to the fact that Xe atoms have the highest energy of interaction with the interior voids of CNTs [16]. The energy of the $N_2$ interaction at these sites is much lower and comparable to the binding energy at the outer surface of CNTs [17].

Previously, high concentrations of the sorbed gas were also observed in the $H_2$—c-SWNT system [4] but the spatial distribution of the $H_2$ molecules in the SWNT bundles was evidently different. The $H_2$ molecules occupied the grooves at the bundle surface, the rest of the bundle surface and the interstitial channels. They also penetrated inside the CNTs with closed ends through the interstitial channels and the surface defects of the tubes. Since the interaction between the $H_2$ molecules was rather weak, they most likely formed a single layer at the bundle surface.

## 2. Radial thermal expansion of $N_2$-saturated SWNT bundles

### Experimental technique

The sample for investigation was prepared by layer-by-layer compression (P = 1.1 GPa [18]) of an SWNT powder. The procedure applied aligned the axes of the CNTs in the plane perpendicular to the sample axis, which was confirmed by an X-ray analysis (see [2] for details). The sample (a cylinder ~7.2 mm high and ~10 mm in diameter) was used to investigate the effect of nitrogen on the radial thermal expansion $\alpha_r$ of SWNT bundles. (Previously this sample was used to measure $\alpha_r$ of Xe-c-SWNT [1] and $H_2$-c-SWNT [2] systems.) The measurements were made in the interval 2.2 – 43 K using a low temperature capacitance dilatometer with a sensitivity of 0.02 nm [19]. Prior to measurement, the cell with the sample of pressure-oriented SWNTs was evacuated at room temperature for 72 hours to remove possible gas impurities. To dope the CNTs, the cell was filled with a $N_2$ gas to P = 760 Torr and held under this condition for about 24 hours until the measurement was started. Then the measuring cell with the sample in the $N_2$ atmosphere was cooled down to 46 K and the saturation with $N_2$ was continued. The $N_2$ gas was added in small portions up to P = 0.1 Torr in the cell. Care was taken to maintain the pressure in the measuring cell no higher that the equilibrium vapor pressure of $N_2$ at this temperature (0.46 Torr [5]), which excluded $N_2$ condensation on the sample surface and the cell components.

After completing the saturation, the measuring cell was cooled down to liquid helium temperature. The thermal expansion was measured in vacuum down to $1 \cdot 10^{-5}$ Torr.

### Results and discussion.

The temperature dependence of the radial thermal expansion coefficient measured on a $N_2$-o-SWNT sample in the interval 2.2 – 43K is shown in Figs. 3a,b (curve 1). Note that the values for $\alpha_r$ obtained for the $N_2$-SWNT system are equilibrium values, since the results measured at each temperature on heating and cooling coincided within the experimental error.

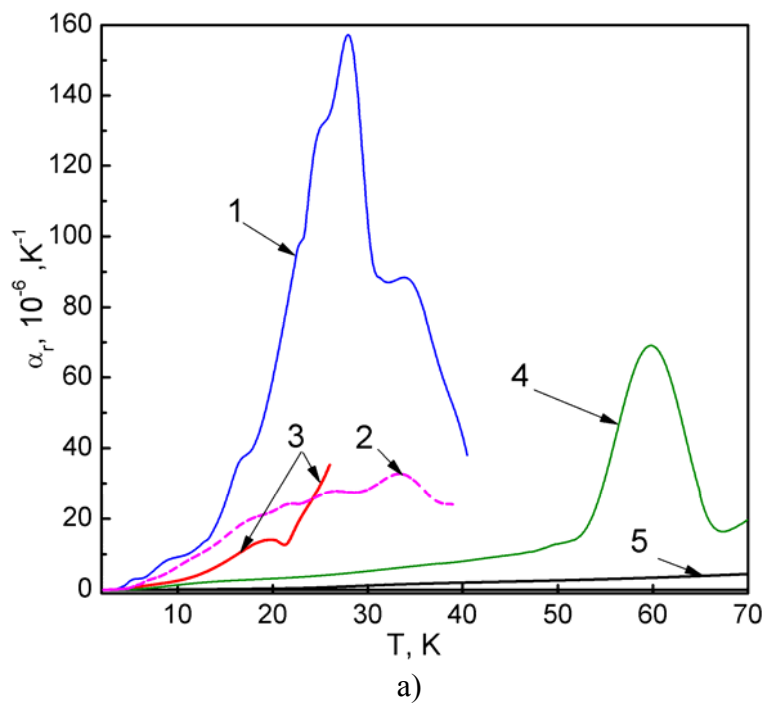

a)

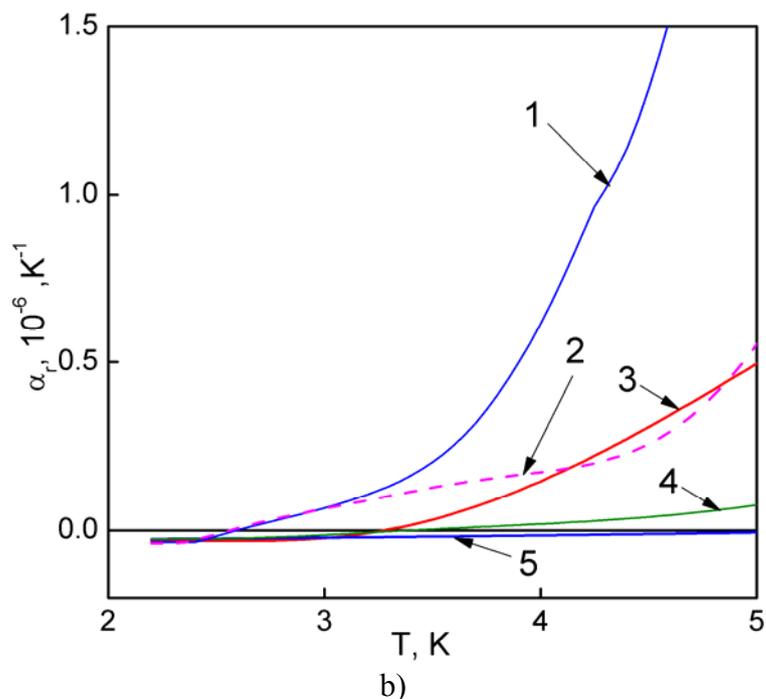

b)

Fig.3. Radial TECs of SWNT bundles: $N_2$-saturated (1);
after a partial $N_2$ removal at T=62K (2);
$H_2$ – saturated (3); Xe – saturated (4);
after a partial Xe removal at T=110K[2] (5);
pure SWNT bundles (6) at T=2.2 – 43 K (a) and 2.2 -5K (b).

It is seen that the magnitudes of $\alpha_r$ increase drastically in the $N_2$-saturated SWNT bundles. The main reason may be that the $N_2$ molecules residing at the lateral surface of the CNTs forming the bundle suppress considerably the transverse vibrations making a negative contribution to the radial thermal expansion [3], which leads to an appreciable increase in $\alpha_r$. A similar effect on the thermal expansion is produced by the $N_2$ molecules sitting in the grooves at the bundle surface (site G), in the interior voids of some CNTs that are at the bundle surface (site $I_{out}$) and in the relatively large interstitial channels (site IC) formed between the different diameter tubes inside the bundles.

At the $N_2$ concentrations used, the $N_2$ molecules are not isolated in the sample. They actually form various $N_2$ aggregations, or clusters. Such clusters make a positive contribution to the thermal expansion of the system. A rough estimate assuming equal TECs for solid $N_2$ and $N_2$ clusters [20] shows that the contribution of the latter does not determine the thermal expansion of the system.

It is interesting that the temperature dependence $\alpha_r$ (T) of the $N_2$-doped SWNT bundles exhibits a very high and wide maximum in the vicinity of T=28K (see Fig.3a curve 1). A similar feature of $\alpha_r$(T) was also found for the Xe-SWNT system (Fig.3a, curve 4). This may be caused by the spatial redistribution of the $N_2$ molecules in the SWNT bundles [21], which involves primarily the $N_2$ molecules that are weakly bound to the SWNT bundles.

We tried to investigate the dependence $\alpha_r$ (T) after removing the weakest – bound $N_2$ molecules from the $N_2$-SWNT system. For this purpose the sample was heated to the temperature 62K which activates (see Section 1 and Fig.3a) the $N_2$ desorption from the bundle surface and possibly from the interior voids of the CNTs with surface defects. The sample was held at T=62K until the $N_2$ desorption was completed and the pressure P$\approx 1 \cdot 10^{-5}$ Torr was reached in the measuring cell. Then it was cooled down to T=2.2K and the thermal expansion was measured again (Fig.3, curve 2). It is seen that the magnitudes of $\alpha_r$ decreased considerably after the removal of $N_2$ from the bundle surface and the interior voids of the CNTs. This suggests that the $N_2$ molecules localized at these sites have a dominant effect on the thermal expansion of the SWNT sample. The contribution from the $N_2$ molecules residing at the sites with high – energy bonds to the SWNT bundles, i.e. in the grooves where they form a quasi-one-dimensional phase, is of secondary importance.

Of interest is also the second $\alpha_r$(T) maximum in the interval 32 – 35K (see Fig.3, curves 1,2). It is reasonable to assume that the $N_2$ molecules adjacent to the CNT surface (Fig.2, layer S) and those forming the group $I_{out}$ interact with the CNTs more strongly than the rest of the molecules, for example, those in the layers S2 and S3 at the bundle surface or inside the CNTs (when they do not contact the CNT walls). The second, higher-temperature maximum in $\alpha_r$(T) accounts for the contribution of the $N_2$ molecules that are in a direct contact with the CNT surface (layer S). This assumption is supported by the almost complete disappearance of the first low temperature maximum of $\alpha_r$(T) after a partial $N_2$ desorption which removes first of all the weakest – bound $N_2$ molecules (layers S2,S3) responsible for its existence.

It should be noted that the partial removal of the $N_2$ impurity (cf. curves 1 and 2 in Fig.3) has a more significant effect on the thermal expansion of the $N_2$-SWNT sample than a similar treatment had on that of the Xe-SWNT system. There may be two reasons for this. Firstly, the impurity concentration is almost an order of magnitude lower in the Xe-doped sample than in the case of $N_2$. Secondly, on partial desorption the $N_2$ molecules are removed from the bundle surface and the interiors of a part of the CNTs while the Xe atoms are cleared only from the outer surface of the SWNT bundles [2].

## Conclusions.

The temperature dependence of the radial thermal expansion coefficient $\alpha_r$(T) of $N_2$ – saturated SWNT bundles has been measured in the interval 2.2 – 43K by the dilatometric method. In the $N_2$-saturated SWNT bundles the magnitudes of $\alpha_r$ increase sharply in the whole range of temperatures used. This is presumably because the $N_2$ molecules suppress the negative contribution to the thermal expansion made by the transverse acoustic vibrations perpendicular to the CNT surface.

The dependence $\alpha_r$(T) of the $N_2$-saturated SWNT bundles has a very wide and high maximum at T~28K which may be caused by the spatial redistribution of the $N_2$ molecules at the bundle surface and inside some CNTs. After a partial $N_2$ desorption at T=62K, lowering the impurity concentration, the magnitudes of $\alpha_r$ decrease sharply in the whole range of temperatures used. Reasons are advanced that such desorption removes the $N_2$ molecules having the weakest binding to the SWNT bundles, i.e. from the sites at the outer CNT surface and inside a part of the CNTs. This means that the determining contribution to the radial thermal expansion of the $N_2$-SWNT system is made by precisely these $N_2$ molecules.

The effects of sorbed $N_2$ and Xe impurities upon the radial thermal expansion of SWNT bundles have been compared qualitatively.

The $N_2$ sorption from a powder consisting of bundles of single-walled carbon nanotubes with open and closed ends has been investigated in the temperature interval 46 – 120K. The air – oxidating treatment of SWNT bundles applied to open the CNT ends enhances the sorptive capacity of the sample by 4.1 times in comparison with the starting SWNT powder.

The authors are indebted to V.M.Loktev, Full Member of the NAS of Ukraine, for fruitful discussions and to the Science & Technology Center of Ukraine (STCU) for the financial support of the study (Project # 4266).